\documentstyle[aps,epsf,preprint]{revtex}

\renewcommand{\slash}[1]{#1 \hspace{-0.55em} / }

\def\bq{\begin{eqnarray}}
\def\eq{\end{eqnarray}}
\def\be{\begin{eqnarray}}
\def\ee{\end{eqnarray}}
\def\ben{\begin{enumerate}}\def\een{\end{enumerate}}

\def\roughly#1{\mathrel{\raise.3ex\hbox{$#1$\kern-.75em
\lower1ex\hbox{$\sim$}}}}

\tighten

\begin{document}


\def\bra{\langle }
\def\ket{\rangle }

\title{
Generalized Parton Distributions
of $^3He$
}

\author{
S. Scopetta}
\address{
Dipartimento di Fisica, Universit\`a degli Studi
di Perugia, via A. Pascoli
06100 Perugia, Italy
\\
and INFN, sezione di Perugia}

\maketitle

\begin{abstract}

A realistic microscopic calculation
of the unpolarized quark Generalized Parton Distribution
(GPD) $H_q^3$
of the $^3He$ nucleus is presented.
In Impulse Approximation, 
$H_q^3$ is obtained as
a convolution between the GPD of the internal nucleon and
the non-diagonal spectral function,
describing properly
Fermi motion 
and binding effects.
The proposed scheme is valid at low values of $\Delta^2$,
the momentum transfer to the target, the most relevant
kinematical region
for the coherent channel of hard exclusive
processes.
The obtained formula has the correct
forward limit, corresponding to the standard
deep inelastic nuclear parton distributions,
and first moment, giving the charge form factor
of $^3He$.
Nuclear effects, evaluated by a
modern realistic potential, 
are found to be larger than in the forward case.
In particular, they
increase with increasing the momentum
transfer when the asymmetry of the process is kept fixed,
and they increase with the asymmetry at fixed
momentum transfer.
Another relevant feature of the obtained
results is that the 
nuclear GPD cannot be factorized into a $\Delta^2$-dependent
and a $\Delta^2$-independent term, as suggested
in prescriptions proposed for finite nuclei.
The size of nuclear effects
reaches 8 \% even 
in the most important part of the kinematical range under scrutiny.
The relevance of the obtained results to study 
the feasibility of experiments is addressed.

\end{abstract}

\section{Introduction}

Generalized Parton Distributions (GPDs) \cite{first} 
parametrize the non-perturbative hadron structure
in hard exclusive 
processes
(for comprehensive reviews, 
see, e.g.,  \cite{jig,rag,pog,dpr}).
The measurement of
GPDs would provide information which is
usually encoded in both the elastic form factors
and 
the usual Parton Distribution Functions (PDFs)
and, 
at the same time, it would represent 
a unique way to access several 
crucial features of the nucleon
\cite{radnew,ji1}, 
such as its angular momentum content 
\cite{ji1}
and its structure in the transverse plane \cite{burk}.
According to a factorization theorem
derived in QCD \cite{fact}, GPDs enter
the long-distance dominated part of
exclusive lepton Deep Inelastic Scattering
(DIS) off hadrons.
In particular, Deeply Virtual Compton Scattering (DVCS),
i.e. the process
$
e H \longrightarrow e' H' \gamma
$ when
$Q^2 \gg m_H^2$,
is one of the the most promising to access GPDs
(here and in the following,
$Q^2$ is the momentum transfer between the leptons $e$ and $e'$,
and $\Delta^2$ the one between the hadrons $H$ and $H'$)
\cite{radnew,ji1,gui}.
Therefore,
relevant experimental efforts to measure GPDs
by means of DVCS off hadrons are likely to
take place in the next few years.
As a matter of fact,
a few DVCS data have been already published 
\cite{hermes,clas}. 

Recently, the issue of measuring GPDs for nuclei
has been addressed. The first paper on this subject
\cite{cano1}, concerning the deuteron,
contained already the crucial observation that
the knowledge of GPDs would permit the investigation
of the short light-like distance structure of nuclei, and thus the interplay
of nucleon and parton degrees of freedom in the nuclear 
wave function.
In standard DIS off a nucleus
with four-momentum $P_A$ and $A$ nucleons of mass $M$,
this information can be accessed in the 
region where
$A x_{Bj} \simeq {Q^2\over 2 M \nu}>1$,
being $x_{Bj}= Q^2/ ( 2 P_A \cdot q )$ and $\nu$
the energy transfer in the laboratory system.
In this region measurements are very difficult, because of 
vanishing cross-sections. As explained in \cite{cano1}, 
the same physics can be accessed
in DVCS at much lower values of $x_{Bj}$.
The usefulness of nuclear GPDs
has been stressed also for finite nuclei in
Ref. \cite{poly}, where it has been shown that
they could provide us with peculiar information
about the spatial distribution of energy, momentum
and forces experienced by quarks and gluons inside hadrons.
Since then, DVCS
has been extensively discussed for
nuclear targets.
Impulse Approximation (IA) calculations,
supposed to give the bulk of nuclear
effects at $0.05 \leq A x_{Bj} \leq 0.7$,
have been performed for
the deuteron \cite{cano2} and for spinless nuclei
\cite{gust}. 
For nuclei of any spin, estimates of GPDs have been
provided and prescriptions for
nuclear effects have been proposed in \cite{km}.
Recently, an analysis of nuclear DVCS
has been performed beyond IA,
with estimates of shadowing
effects and involving therefore large light-like distances and correlations
in nuclei \cite{frest}. It was found that
these effects are sizable up to $A x_{Bj} \simeq 0.1$,
i.e. up to larger values of $A x_{Bj}$ with respect to 
normal DIS. Besides, 
the possibility of measuring DVCS at an 
electron-ion-collider has also been established
in \cite{frest}.

The study of GPDs for $^3He$ is interesting
for many aspects. 
In fact, $^3He$ is a well known nucleus, for which realistic studies 
are possible, so that conventional nuclear effects
can be safely calculated. Strong deviations from the predicted
behaviour could be therefore ascribed to exotic effects, such as
the ones of non-nucleonic degrees of freedom, not included in a
realistic wave function.
Besides, $^3He$ is extensively used as an effective neutron target.
In fact, the properties of the free neutron are being investigated
through experiments with nuclei, whose data
are analyzed taking nuclear effects properly into account.
Recently, it has been shown that unpolarized 
DIS off three body systems can provide relevant information
on PDFs at large $x_{Bj}$ \cite{thom1,io1,sim,rin}, while
it is known since a long time that its particular spin structure
suggests the use of $^3He$ as an effective polarized
neutron target \cite{friar,io2,sauer,gfst}.
Polarized $^3He$ will be therefore the first candidate
for experiments aimed at the study of spin-dependent
GPDs in the free neutron, to unveil details of its angular momentum
content. 

In this paper, an Impulse Approximation (IA)
calculation of the quark unpolarized GPD $H_q^3$ of
$ ^3He $ is presented. A convolution formula
is derived and afterwards numerically evaluated
using a realistic non-diagonal spectral function,
so that Fermi motion and binding effects are rigorously estimated.
The proposed scheme is valid for $\Delta^2 \ll Q^2,M^2$
and despite of this it permits to
calculate GPDs in the kinematical range relevant to
the coherent, no break-up channel of deep exclusive processes off $^3He$.
In fact, the latter channel is the most interesting one for its 
theoretical implications, but it can be hardly observed at
large $\Delta^2$, due to the vanishing cross section
\cite{frest}.
The nuclear GPDs obtained here
are therefore a prerequisite for any calculation of observables 
in coherent DVCS off $^3He$, 
although they can be compared neither with existing
data, nor with forth-coming ones.
A detailed analysis of DVCS off $^3He$, with estimates
of observables, such as
cross-sections or spin asymmetries, is in progress
and will follow in a separate publication.
Thus, the main result of this investigation
is not the size and shape of the obtained $H_q^3$ for $^3He$,
but the size and nature of nuclear effects on it.
This will permit to test directly, for the
$^3He$  target at least, the accuracy of prescriptions
which have been proposed to estimate nuclear GPDs \cite{km},
providing a useful tool for the planning of future experiments
and for their correct interpretation.


The paper is organized as follows:
The notation used is introduced in section 2, together
with the derivation of the main result, a convolution formula
obtained from an IA analysis.
Such a formula is used to evaluate the numerical results
shown in section 3 and thoroughly discussed in section 4,
where the most interesting outcome of the present study is to be found.
Eventually, conclusions are drawn in the fifth section.

\section{Formalism}


The formalism introduced in Ref. \cite{jig} is adopted.
One has to think to a spin $1/2$ hadron target, with initial (final)
momentum and helicity $P(P')$ and $s(s')$, 
respectively. 
The GPDs $H_q(x,\xi,\Delta^2)$ and
$E_q(x,\xi,\Delta^2)$
are defined through the expression
\begin{eqnarray}
\label{eq1}
F^q_{s's}(x,\xi,\Delta^2) & = &
{1 \over 2} \int {d \lambda \over 2 \pi} e^{i \lambda x}
\bra P' s' | \, \bar \psi_q \left(- {\lambda n \over 2}\right)
\slash{n} \, \psi_q \left({\lambda n \over 2} \right) | P s \ket  =  
\nonumber
\\
& = & H_q(x,\xi,\Delta^2) {1 \over 2 }\bar U(P',s') 
\slash{n} U(P,s) + 
E_q(x,\xi,\Delta^2) {1 \over 2} \bar U(P',s') 
{i \sigma^{\mu \nu} n_\mu \Delta_\nu \over 2M} U(P,s)~,
\end{eqnarray}
\\
where 
$\Delta=P^\prime -P$
is the 4-momentum transfer to the hadron,
$\psi_q$ is the quark field and M is the hadron mass.
It is convenient to work in
a system of coordinates where
the photon 4-momentum, $q^\mu=(q_0,\vec q)$, and $\bar P=(P+P')/2$ 
are collinear along $z$.
The $\xi$ variable in the arguments of the GPDs 
is the so called ``skewedness'', parametrizing
the asymmetry of the process. It is defined
by the relation 
\bq
\xi = - {n \cdot \Delta \over 2} = - {\Delta^+ \over 2 \bar P^+}
= { x_{Bj} \over 2 - x_{Bj} } + {\cal{O}} \left ( {\Delta^2 \over Q^2}
\right ) ~,
\label{xidef}
\eq
where $n$
is a light-like 4-vector
satisfying the condition $n \cdot \bar P = 1$.
One should notice that the variable $\xi$
is completely fixed by the external lepton kinematics.
Here and in the following, use is made of the definition
$a^{\pm}=(a^0 \pm a^3)/\sqrt{2}$.
As explained in \cite{ji1},
GPDs describe the amplitude for finding a quark with momentum fraction
$~~x+\xi$ (in the Infinite Momentum Frame) in a hadron 
with momentum $(1+\xi) \bar P$
and replacing it back into
the nucleon with a momentum transfer $\Delta$.
Besides, when the quark longitudinal momentum fraction 
$x$ of the average nucleon momentum $\bar P$
is less than $-\xi$, GPDs describe antiquarks;
when it is larger than $\xi$, they describe quarks;
when it is between $-\xi$ and $\xi$, they describe 
$q \bar q$ pairs.
One should keep in mind that, in addition to the variables
$x,\xi$ and $\Delta^2$ explicitly shown, GPDs depend,
as the standard PDFs, on the momentum scale $Q^2$ at which they are
measured or calculated. 
Such a dependence will not be discussed in the present paper and,
for an easy presentation, it will be 
omitted.
The values of $\xi$ which are possible for a given value of
$\Delta^2$ are:
\bq
0 \le \xi \le \sqrt{- \Delta^2}/\sqrt{4 M^2-\Delta^2}~.
\label{xim}
\eq
The well known natural constraints of $H_q(x,\xi,\Delta^2)$ are: 

i) the so called
``forward'' or ``diagonal'' limit, 
$P^\prime=P$, i.e., $\Delta^2=\xi=0$, where one 
recovers the usual PDFs
\bq
H_q(x,0,0)=q(x)~;
\label{i)}
\eq

ii)
the integration over $x$, yielding the contribution
of the quark of flavour $q$ to the Dirac 
form factor (f.f.) of the target:
\bq
\int dx H_q(x,\xi,\Delta^2) = F_1^q(\Delta^2)~;
\label{ii)}
\eq

iii) the polynomiality property \cite{jig},
involving higher moments of GPDs, according to which
the $x$-integrals of $x^nH^q$ and of $x^nE^q$
are polynomials in $\xi$ of order $n+1$.

In \cite{epj,io3},
an expression
for $H_q(x,\xi,\Delta^2)$ of a given hadron target, 
for small values of $\xi^2$, has been obtained
from the definition Eq. (\ref{eq1}).

It reads:
\bq
H_q(x,\xi,\Delta^2)
& = &
\int 
{d \vec k } \,\,
\delta \left( {k^+ \over \bar P^+} - (x+\xi) \right)
\times
\nonumber
\\
& \times &
\left [ 
{ 1 \over (2 \pi)^3 V}
{k^+ \over k_0}
\sum_\lambda 
\bra P' | 
b_{q,\lambda}^\dag(k^+ + \Delta^+, \vec k_\perp + \vec \Delta_\perp)
b_{q,\lambda}(k^+, \vec k_\perp )
| P \ket \right] + O(\xi^2)~,
\label{int_1}
\eq
where states and creation and annihilation operators 
are normalized as follows
\bq
\bra P' |  P \ket = (2 \pi)^3 
\delta(P'^+ - P^+)
\delta(\vec P'_\perp - \vec P_\perp)~
\label{nors}
\eq
and
\bq
\{ b(k^+,\vec k_\perp),b^\dag (k'^+, \vec k'_\perp)\} = (2 \pi)^3 
\delta(k'^+ - k^+)
\delta(\vec k'_\perp - \vec k_\perp)~,
\label{norb}
\eq
respectively.

All the steps leading from Eq. (\ref{eq1}) to Eq. (\ref{int_1})
are fully described in Ref. \cite{io3}. They include initially the
use of light-cone spinors and fields, neglecting terms of
order $O(\xi^2)$ in the right hand side of Eq. (\ref{eq1}), 
and later a transition
to the expression (\ref{int_1}),
where the same quantities are normalized according to Eqs.
(\ref{nors}) and (\ref{norb}).

The case of a spin $1/2$ 
hadron with three composite constituents
has been discussed in \cite{io3}.
In that paper,
the GPDs of the proton have been studied, assuming that
it is made of complex constituent quarks.
In here, the approach will be extended to $^3He$.
The GPD $H_q^3$ of $^3He$ will be obtained in IA as
a convolution between
the non-diagonal spectral function of the internal nucleons,
and the GPD $H_q^N$ of the nucleons themselves. 

The scenario is depicted in Fig. 1 for the 
special case of coherent DVCS, in the handbag
approximation. One parton
with momentum $k$,
belonging to a given nucleon
of momentum $p$ in the nucleus, interacts with the probe
and it is afterwards reabsorbed, with momentum
$k+\Delta$, by the same nucleon,
without further re-scattering with the recoiling system
of momentum $P_R$. Finally, the interacting nucleon
with momentum $p + \Delta$ is reabsorbed back into the nucleus.
The analysis suggested here
is quite similar to the usual IA approach
to DIS off nuclei \cite{fs,cio}.

In the class of frames discussed above,
and in addition to the kinematical variables,
$x$ and $\xi$, already defined,
one needs a few more to describe the process.
In particular,
$x'$ and $\xi'$, for the ``internal''
target, i.e., the nucleon, have to be introduced.
The latter quantities can be obtained defining the ``+''
components of the momentum $k$ and $k + \Delta$ of the struck parton
before and after the interaction, with respect to
$\bar P^+$ and $\bar p^+ = {1 \over 2} (p + p')^+$:
\begin{eqnarray}
k^+ & = & (x + \xi ) \bar P^+ =  (x' + \xi') \bar p^+~,
\\
(k+\Delta)^+ & = & 
(x - \xi ) \bar P^+ =  (x' - \xi') \bar p^+~.
\end{eqnarray}
From the above expressions, $\xi'$ and $x'$ are immediately obtained as
\begin{eqnarray}
\xi' & = & - { \Delta^+ \over 2 \bar p^+}
\label{defxi1}
\\
x' & = & {\xi' \over \xi} x
\label{defx1}
\end{eqnarray}
and, since $\xi = - \Delta^+ / (2 \bar P^+)$, 
if $\tilde z = p^+/P^+$, one also has
\begin{eqnarray}
\xi ' = {\xi \over \tilde z( 1 + \xi ) - \xi}~.
\label{xi1}
\end{eqnarray}

These expressions have been already found and used
in 
\cite{cano1,gust,km,io3}.

In order to derive a convolution formula in IA for $H_q^3$,
the procedure used for standard DIS will be adopted \cite{fs,cio}.
First, in Eq. (\ref{int_1}),
two complete sets of states, corresponding
to the interacting nucleon and to the recoiling two-body
system, are properly inserted 
to the left and right-hand sides of the quark operator:
\begin{eqnarray}
{H_q^3(x,\xi,\Delta^2)} & = &  
{\bra P'S |} \,\,\sum_{ P_R',S_R', p',s'}
{ \{ | P_R' S_R' \ket | p' s' \ket \}} 
\{ \bra  P_R' S_R' |  {\bra  p' s'|} \} 
\nonumber
\\
& &
\int 
{d \vec k \over (2 \pi )^3 }
{ k^+ \over k_0} 
\delta \left( {k^+ \over \bar P^+} - (x+\xi) \right)
{ 1 \over V}
\sum_{q,\lambda} 
b_{q,\lambda}^\dag(k^+ + \Delta^+, \vec k_\perp + \vec \Delta_\perp)
b_{q,\lambda}(k^+, \vec k_\perp )
\nonumber
\\
& &
\sum_{ P_R,S_R,p,s}
\{ |  P_R S_R \ket {| p s \ket \}} 
{\{ \bra  P_R S_R |  \bra  p s| \} 
\,\,\,\,|   P S  \ket}~;
\label{iaint}
\end{eqnarray}
second, 
since eventually use has to be made of NR nuclear wave functions, a 
state $| \vec p \ket $
has to be normalized in a NR manner:
\bq
\bra \vec p' | \vec  p \ket = (2 \pi)^3 
\delta(\vec p' - \vec p)~,
\label{norr}
\eq
so that in Eq. (\ref{iaint}) one has to perform
the substitution:
\bq
| p > \rightarrow \sqrt{{p^+ \over p^0} }| \vec p >~. 
\eq
Since, using IA in the intrinsic frame of $^3He$ one has,
for the NR states:
\begin{eqnarray}
{\{ \bra  \vec P_R S_R |  \bra  \vec p s| \} 
| \vec P S \ket} = {\bra \vec P_R S_R, \vec p s  
| \vec P S \ket } (2 \pi)^3 \delta^3 (\vec P - \vec P_R - \vec p)
\delta_{S,S_R\,s}~,
\nonumber
\end{eqnarray}
a convolution formula is readily obtained from Eq. (\ref{iaint}):
\begin{eqnarray}
H_q^3(x,\xi,\Delta^2) & \simeq & 
\sum_N \int dE \int d \vec p
\, 
\sqrt{p^+ (p+\Delta)^+ \over p_0 (p + \Delta)_0}
P_{N}^3(\vec p, \vec p + \vec \Delta, E ) 
{\xi' \over \xi}
H_{q}^N(x',\xi',\Delta^2) + 
{\cal{O}} 
\left ( \xi^2 \right ) =
\label{flux}
\\
& =  & 
\sum_N \int dE \int d \vec p
\, 
[ P_{N}^3(\vec p, \vec p + \vec \Delta, E ) + 
{\cal{O}} 
( {\vec p^2 / M^2},{\vec \Delta^2 / M^2}) ]
\nonumber
\\
& \times & 
{\xi' \over \xi}
H_{q}^N(x',\xi',\Delta^2) + 
{\cal{O}} 
\left ( \xi^2 \right )~.
\label{spec}
\end{eqnarray}
In the above equations, the kinetic energies of the residual nuclear
system and of the recoiling nucleus have been neglected, and
$P_{N}^3 (\vec p, \vec p + \vec \Delta, E )$ is
the one-body off-diagonal spectral function
for the nucleon $N$ in $^3He$:
\begin{eqnarray}
P_N^3(\vec p, \vec p + \vec \Delta, E)  & = & 
{1 \over (2 \pi)^3} {1 \over 2} \sum_M 
\sum_{R,s}
\bra \vec P'M | (\vec P - \vec p) S_R, (\vec p + \vec \Delta) s\ket 
\bra (\vec P - \vec p) S_R,  \vec p s| \vec P M \ket
\times
\nonumber
\\
& \times &  
\, \delta(E - E_{min} - E^*_R)~.
\label{spectral}
\end{eqnarray}

Besides, the quantity

\bq
H_q^N(x',\xi',\Delta^2)
& = &
\int 
{d \vec k } \,\,
\delta \left( {k^+ \over \bar p^+} - (x'+\xi') \right)
\nonumber
\\
& \times &
{ 1 \over (2 \pi)^3 V}
{k^+ \over k_0}
\sum_\lambda 
\bra p' | 
b_{q,\lambda}^\dag(k^+ + \Delta^+, \vec k_\perp + \vec \Delta_\perp)
b_{q,\lambda}(k^+, \vec k_\perp ) | p \ket
=
\nonumber
\\
& = &
{\xi \over \xi'}
\int 
{d \vec k } \,\,
\delta \left( 
{k^+ \over \bar P^+} - 
(x +\xi) \right)
\nonumber
\\
& \times &
{ 1 \over (2 \pi)^3 V}
{k^+ \over k_0}
\sum_\lambda 
\bra p' | 
b_{q,\lambda}^\dag(k^+ + \Delta^+, \vec k_\perp + \vec \Delta_\perp)
b_{q,\lambda}(k^+, \vec k_\perp ) | p \ket
\label{gpdb}
\eq
is, according to Eq. (\ref{int_1}),
the GPD of the bound nucleon N
up to terms of order $O(\xi^2)$, and in the above equation
use has been made
of Eqs. (\ref{defxi1}) and (\ref{defx1}).

The delta function in Eq (\ref{spec})
defines $E$, the so called removal energy, in terms of
$E_{min}=| E_{^3He}| - | E_{^2H}| = 5.5$ MeV and
$E^*_R$, the excitation energy 
of the two-body recoiling system.
The main quantity appearing in the definition
Eq. (\ref{spectral}) is
the overlap integral
\bq
\bra \vec P M | \vec P_R S_R, \vec p s \ket=
\int d \vec y \, e^{i \vec p \cdot \vec y}
\bra \chi^{s},
\Psi_R^{S_R}(\vec x) | \Psi_3^M(\vec x, \vec y) \ket~,
\label{trueover}
\eq 
between the eigenfunction 
$\Psi_3^M$ 
of the ground state
of $^3He$, with eigenvalue $E_{^3He}$ and third component of
the total angular momentum $M$, and the
eigenfunction $\Psi_R^{S_R}$, with eigenvalue
$E_R = E_2+E_R^*$ of the state $R$ of the intrinsic
Hamiltonian pertaining to the system of two interacting
nucleons \cite{over}.
Since the set of the states $R$ also includes
continuum states of the recoiling system, the summation
over $R$ involves the deuteron channel and the integral
over the continuum states.

Concerning Eqs (\ref{spec}--\ref{trueover}), two comments are in order.

The first
concerns the accuracy of the actual calculations
which will be presented. In the following, a NR spectral function
will be used to evaluate Eq. (\ref{spec}), 
so that the accuracy of the calculation is
of order 
${\cal{O}} 
\left ( {\vec p^2 / M^2},{\vec \Delta^2 / M^2} \right )$.
In fact, if use is made of NR wave 
functions to calculate the non-diagonal spectral function,
the result holds for 
$\vec p^2, ( \vec p + \vec \Delta)^2 << M^2$. 
The same constraint can be written
$\vec p^2, \vec \Delta^2 << M^2$.
While the first of these conditions is the usual one
for the NR treatment of nuclei, the second
forces one to use Eq. (\ref{spec}) only
at low values of $\vec \Delta^2$, for which the accuracy is good enough.
As it will be explained, the interest of the present
calculation is indeed to investigate nuclear effects at low values
of $\vec \Delta^2$.

Second, from the study of forward DIS off nuclei, it is known
that, to properly estimate nuclear effects,
in going from a covariant formalism
to one where use can be made of the usual
nuclear wave functions, one has to 
keep the correct normalization
\cite{fs,ku}. This procedure
leads to the appearance of the ``flux factor'',
represented in Eq. (\ref{flux}) by the expression
$\sqrt{ p^+ (p + \Delta)^+ / p_0 (p + \Delta)_0 }$
(which gives, in forward DIS, the usual $p^+ / p_0$ term).
This factor gives one at order
${\cal{O}} 
\left ( {\vec p^2 / M^2},{\vec \Delta^2 / M^2} \right )$,
which is the accuracy of the present analysis, and 
it will not be included in the actual
calculation. 
In other words, if one takes into account
that the intrinsic light-front frame and the usual rest frame
of $^3He$, where the wave functions are evaluated, are not
the same frame, the flux factor
should be introduced \cite{fs,ku}; anyway,
in the NR approach used here, the two frames do not differ.  
This is an effect of the used approximations, which have to be
relaxed if one wants to be predictive in more general processes
at higher momentum transfer. In fact,
in that case the accuracy of Eq. (\ref{spec}) is not good 
enough any more
and the calculation, in order to be consistent, has to be
performed taking into account relativistic 
corrections. Work is being done presently in this direction, and
the discussion of this important point will be included in
a following paper, describing
a calculation performed in a light-front framework.
One final remark about the flux-factor: to neglect the corresponding
quantity for the quark inside the internal nucleon,
i.e. the term $k^+/k_0$ in the GPD of the bound nucleon, Eq. (\ref{gpdb}),
would be a rough approximation, with respect to the
nuclear case, because the motion of quarks in the nucleon
is relativistic,
even at the constituent level.
In any case in the present calculation, as described
in the following section, the GPD for the bound quark
will not be evaluated, while an available model for it 
will be used.
The main emphasis of the present approach, as already said,
is not on the absolute values of the results, but in the nuclear effects,
which can be estimated by taking any reasonable form for
the internal GPD.

Eq. (\ref{spec}) can be written in the form
\begin{eqnarray}
H_{q}^3(x,\xi,\Delta^2) & = & 
\sum_N \int d E
\int d \vec p
\, P_N^3(\vec p, \vec p + \vec \Delta) {\xi' \over \xi}
H_q^N(x',\xi',\Delta^2) = \nonumber
\\
& = &  
\sum_N \int_x^1 {dz \over z} 
\int d E
\int d \vec p
\, P_N^3(\vec p, \vec p + \vec \Delta) 
\delta \left( z - { \xi \over \xi' }  \right)
H_q^N \left( { x \over z }, { \xi \over z },\Delta^2 \right)~.
\label{pnk}
\end{eqnarray}
Taking into account that 
\begin{equation}
z - { \xi \over \xi'} = z - [ \tilde z ( 1 + \xi  ) - \xi ]
= z + \xi - { p^+ \over P^+} ( 1 + \xi )
= z + \xi  - { p^+ \over \bar P^+}~,
\end{equation}
Eq. (\ref{pnk}) can also be written in the form:
\begin{eqnarray}
H_{q}^3(x,\xi,\Delta^2) =  
\sum_N \int_x^1 { dz \over z}
h_N^3(z, \xi ,\Delta^2 ) 
H_q^N \left( {x \over z},
{\xi \over z},\Delta^2 \right)~,
\label{main}
\end{eqnarray}
where 
\begin{equation}
h_N^3(z, \xi ,\Delta^2 ) =  
\int d E
\int d \vec p
\, P_N^3(\vec p, \vec p + \vec \Delta) 
\delta \left( z + \xi  - { p^+ \over \bar P^+ } \right)~.
\label{hq0}
\end{equation}

One should notice that
Eqs. (\ref{main}) and (\ref{hq0}) or, which is the same,
Eq. (\ref{spec}), fulfill the constraint $i)-iii)$ previously listed.

The constraint $i)$, i.e. the forward limit
of GPDs, is certainly verified.
In fact, by taking
the forward limit ($\Delta^2 \rightarrow 0, \xi \rightarrow 0$)
of Eq. (\ref{main}), one gets the 
expression which is usually found,
for the parton distribution $q_3(x)$, in the IA analysis of
unpolarized DIS off $^3He$ 
\cite{io1,fs,cio}: 
\begin{eqnarray}
q_3(x) =  H_q^3(x,0,0) =
\sum_{N} \int_x^1 { dz \over z}
f_{N}^3(z) \,
q_{N}\left( {x \over z}\right)~.
\label{mainf}
\end{eqnarray}
In the latter equation,
\begin{equation}
f_{N}^3(z) = h_{N}^3(z, 0 ,0) =  \int d E \int d \vec p
\, P_{N}^3(\vec p,E) 
\delta\left( z - { p^+ \over \bar P^+ } \right)
\label{hq0f}
\end{equation}
is the light-cone momentum distribution of the nucleon $N$
in the nucleus, $q_N(x)= H_q^N( x , 0, 0)$
is the distribution
of the quark of flavour $q$ 
in the nucleon $N$ and $P_N^3(\vec p, E)$,
the $\Delta^2 \longrightarrow 0$ limit of
Eq. (\ref{pnk}), is the
one body spectral function.

The constraint $ii)$, i.e. the $x-$integral of the GPD
$H_q$, is also naturally
fulfilled. In fact, by $x-$integrating Eq. (\ref{main}),
one easily obtains:
\begin{eqnarray}
\int dx H_q^3(x,\xi,\Delta^2) & = & \sum_N
\int dx \int {dz \over z} h_N^3(z,\xi,\Delta^2)
H_q^N \left ( { x \over z}, {\xi \over z},
\Delta^2 \right ) =
\nonumber
\\
& = &
\sum_N
\int d x' H_q^N (x',\xi',\Delta^2) \int d z 
h_N^3(z,\xi,\Delta^2) =
\nonumber
\\
& = &
\sum_N 
F_q^N(\Delta^2)
F_N^3(\Delta^2)
= F_q^3(\Delta^2)~.
\label{ffc}
\eq
In the equation above,
$F_q^3(\Delta^2)$ is the
contribution, 
of the quark of flavour $q$,
to the
nuclear f.f.;
$F_q^N(\Delta^2)$ is the contribution,
of the quark of flavour $q$,  
to the nucleon $N$ f.f.;
$F_N^3(\Delta^2)$ is the 
so-called $^3He$ ``pointlike f.f.'', which
would represent the contribution of the nucleon $N$ to the
f.f. of $^3He$ if $N$ were point-like.
$F_N^3(\Delta^2)$ is given, in the present approximation, by
\bq
F_N^3(\Delta^2) = \int dE \int d \vec p
\, P_N^3(\vec p, \vec p + \vec \Delta, E)
= \int dz \, h_N^3(z,\xi,\Delta^2)~. 
\label{ffp}
\eq

Eventually the polynomiality, condition $iii)$,
is formally fulfilled by Eq. (\ref{spec}), although
one should always remember that it is a 
result of order ${\cal{O}}(\xi^2)$,
so that high moments cannot be really checked.

Summarizing this section, we have derived in IA a convolution
formula for the GPD $H_q^3$ of $^3He$, Eq. (\ref{spec}) (or,
which is equivalent, Eq. (\ref{main})), at order
${\cal{O}}\left ({\vec p^2 \over M^2}, 
{\vec \Delta^2 \over M^2}, \xi^2 \right )$, 
in terms of a non-diagonal nuclear one-body
spectral function, Eq. (\ref{spectral}), and of the GPD $H_q^N$
of the internal nucleon.

\section{Numerical Results}

$H_q^3(x,\xi,\Delta^2)$, Eq. (\ref{spec}), 
has been evaluated in the nuclear Breit Frame.

The non-diagonal spectral function
Eq. (\ref{spectral}), appearing in Eq.
(\ref{spec}),
has been calculated 
along the lines of Ref. \cite{gema},
by means of 
the overlap Eq. (\ref{trueover}), which 
exactly includes
the final state interactions in the two nucleon recoiling system,
the only plane wave being that describing the relative motion
between the knocked-out nucleon and the two-body system
\cite{over}. 
The realistic wave functions $\Psi_3^M$
and $\Psi_R^{S_R}$ in Eq. (\ref{trueover})
have been evaluated
using the 
AV18 interaction \cite{av18} 
and
taking into account
the Coulomb repulsion of protons in $^3He$.
In particular $\Psi_3^M$ has been 
developed along the lines of Ref. \cite{tre}.
The same overlaps have been already used in Ref.\cite{io1}.

The other ingredient in Eq. (\ref{spec}), i.e.
the nucleon GPD $H_q^N$, has been modelled in agreement with
the Double Distribution representation \cite{radd}, as described
in \cite{rad1}. 
For the reader convenience,
the explicit form of $H_q^N$ is listed below \cite{rag,radd}:
\begin{eqnarray}
H_q^N(x,\xi,\Delta^2) = \int_{-1}^1 d\tilde x
\int_{-1 + |\tilde x|}^{1-|\tilde x|} 
\delta(\tilde x + \xi  \alpha - x)
\tilde \Phi_{q} (\tilde x, \alpha,\Delta^2) d \alpha~.
\label{hdd}
\end{eqnarray}

With some care, the expression above can be integrated over 
$\tilde x$ and the result is explicitly given in 
\cite{rag}.

In \cite{radd}, a factorized
ansatz is suggested for the DD's: 
\begin{equation}
\tilde \Phi_{q} (\tilde x, \alpha,\Delta^2) =
h_{q} (\tilde x, \alpha,\Delta^2)
\Phi_{q} (\tilde x) 
F_{q}(\Delta^2)~,
\label{ans}
\end{equation}
with 
the $\alpha$ dependent
term, 
$h_{q} (\tilde x, \alpha,\Delta^2)$, 
which has the character of a mesonic amplitude,
fulfilling the relation:
\begin{equation}
\int_{-1 + |\tilde x|}^{1-|\tilde x|} h_{q} 
(\tilde x, \alpha,\Delta^2) d \alpha = 1~.
\label{hnor}
\end{equation}
Besides, in Eq. (\ref{ans}) 
$
\Phi_{q} (\tilde x) 
$
represents the forward density
and, eventually, 
$
F_{q}(\Delta^2)
$
the contribution of the quark of flavour $q$
to the nucleon form factor.
Eq. (\ref{hdd}) fulfills the
crucial constraints of GPDs, i.e., the forward limit,
the first-moment and the polynomiality condition.
One needs now to model the three functions
appearing in Eq. (\ref{ans}).
For the amplitude $h_q$, use will be made of
one of the simple normalized forms suggested in
\cite{radd}, 
on the bases of the symmetry
properties of DD's:
\begin{equation}
h_{q}^{(1)}(\tilde x, \alpha) = {3 \over 4} { (1 - \tilde x)^2 - \alpha^2 
\over (1 - \tilde x)^3 }~;
\label{h}
\end{equation}

For the forward distribution $\Phi_q(\tilde x)$, the general simple form
\begin{equation}
\Phi_q(\tilde x) = { \Gamma(5-a) \over 6 \Gamma(1-a)} \tilde x^{-a} (1-
\tilde x)^3~,
\label{for}
\end{equation}
with $a=0.5$, has been taken.

Concerning the model used for the internal nucleon $H_q^N$,
two caveats are in order.

First of all, Eq. (\ref{hdd}), with the choices Eqs. (\ref{h}) 
and (\ref{for}),
corresponds to the
$NS$, valence quark contribution proposed in \cite{rad1}, which is
symmetric in the variable $x$.
I stress again that the main point of the present study
is not to produce realistic estimates for observables,
but to investigate and discuss nuclear effects,
which do not depend on the form of any well-behaved internal
GPD, whose general structure is safely simulated by Eqs.
(\ref{hdd}) -- (\ref{for}).
For the moment being, only the $NS$ part of
$H_q^N$ is therefore calculated; if one wanted to estimate
the DVCS cross-sections, also the quark
singlet and the gluon contributions
would be certainly necessary. As already said, a detailed study of DVCS
on $^3He$ is in progress and will be shown elsewhere.

Secondly, to take advantage of the used 
parametrization, Eqs. (\ref{hdd}) -- (\ref{for}),
of the nucleonic GPD, one should work in a symmetric
frame for the nucleon, such as the Breit Frame. 
Actually, the present calculation is performed
in the nuclear Breit Frame, which does not coincide 
with the Breit Frame of the internal nucleon.
For this reason, one should perform a boost to properly work
in the nuclear Breit Frame.
Anyway, as it is pointed out in Ref. \cite{gust},
any NR evaluation of nuclear
GPDs, as a result of the NR reduction,
is intrinsically frame dependent
and it is valid only with accuracy 
${\cal O}(p^2/M^2)$, like Eq. (\ref{spec}).
In this calculation therefore, 
the effect of using a different 
reference frame for the nucleon is not taken
into account, since
in any case it would be an effect of order
${\cal O}(p^2/M^2)$, which is the overall accuracy of the present
work [cf. Eq. (\ref{spec})]. In the above procedure, therefore,
no further approximations are introduced
in addition to the ones leading to Eq. (\ref{spec}).
To overcome such a problem, one could use a light-front
approach \cite{kp}, as it is done for the GPDs of the
deuteron in \cite{cano1,cano2}. 

Eventually,
the $F_q(\Delta^2)$ term in Eq. (\ref{ans}), i.e. the contribution
of the quark of flavour $q$
to the nucleon form factor, has been obtained from
the experimental values of the proton, $F_1^p$, and
of the neutron, $F_1^n$, Dirac form factors. For the
$u$ and $d$ flavours, neglecting the effect of the 
strange quarks, one has
\bq
F_u (\Delta^2)& = & {1 \over 2} (2 F_1^p(\Delta^2) + F_1^n(\Delta^2))~,
\nonumber
\\
F_d (\Delta^2)& = & 2 F_1^n(\Delta^2) + F_1^p(\Delta^2)~.
\label{fq}
\eq
The contributions of the flavours $u$ and $d$
to the proton and neutron f.f. are therefore
\bq
F_u^p (\Delta^2)& = & {4 \over 3} F_u(\Delta^2)~,
\nonumber
\\
F_d^p & = & - {1 \over 3} F_d(\Delta^2) 
\label{fpq}
\eq
and
\bq
F_u^n (\Delta^2)& = & 
{2 \over 3} F_d(\Delta^2)~,
\nonumber
\\
F_d^n (\Delta^2) & = & - {2 \over 3} F_u(\Delta^2)~,
\label{fnq}
\eq 
respectively.

For the numerical calculations,
use has been made of the parametrization of the nucleon
Dirac f.f. given in Ref. \cite{gari}.

Now
the ingredients of the calculation 
have been completely described, so that numerical results
can be presented.
First of all, the forward limit of $H_q^3$, Eq.
(\ref{mainf}), will be discussed, together with its $x-$integral,
Eq. (\ref{ffc}).

In Fig. 2, it is shown the forward limit of the ratio
\bq
R_q (x,\xi,\Delta^2) = 
{ H_q^3(x,\xi,\Delta^2) 
\over 
2 H_q^p(x,\xi,\Delta^2) + H_q^n(x,\xi,\Delta^2)}~,
\label{rat}
\eq
i.e. the quantity
\bq
R_q (x,0,0) = 
{ H_q^3(x,0,0) \over 2 H_q^p(x,0,0) + H_q^n(x,0,0)}
=  { q^3(x) \over 2 q^p(x) + q^n(x)} 
\label{rat1}
\eq
for the flavour $q=u,d$, as a function of $x_3=3x$.
In the above equation, the numerator is given by Eq. (\ref{mainf}),
while the denominator clearly represents
the distribution of the
quarks of flavour $q$ 
in $^3He$ if nuclear effects are completely
disregarded, i.e., the interacting quarks are assumed to belong
to free nucleons at rest.
The behaviour which is found is typically $EMC-$like,
as it is usually obtained in IA studies of DIS
on nuclei \cite{io1,fs,cio}, so that, 
in the forward limit, well-known results are recovered.
One should notice that,
had the structure functions ratio, instead of the PDFs one,
been shown, the ratio would be 1 for $x=0$, due to the 
normalization of the
spectral function in Eq. (\ref{hq0f}).
It is also useful, for later convenience, to realize
that nuclear effects for the $d$ flavour are a bit larger
than those which are found for the $u$ flavour.
This is due to the fact that the forward $d$ distribution
is more sensitive than the $u$ one to the neutron
light-cone distribution, Eq. (\ref{hq0f}), which
is different from that of the proton.
In fact, the average momentum of the neutron
in $^3He$ is larger than the one of the proton
\cite{over,cio}. 

In Fig. 3, the quantity
\bq
{ 1 \over 2 } \sum_q \int dx H_q^3(x,\xi,\Delta^2) =  
{ 1 \over 2 } \sum_q \sum_N
\int dx \int {dz \over z} h_N^3(z,\xi,\Delta^2)
H_q^N \left ( { x \over z}, {\xi \over z},
\Delta^2 \right )~,
\eq
is shown.
According to Eq. (\ref{ffc}),
in the present approximation,
this should give, 
up to 
contributions of heavier quarks,
the charge form factor 
$F_{ch}^3(\Delta^2)$
of $^3He$ (the usual normalization
$F_{ch}^3(0)=1$ is chosen): 
\bq
F_{ch}^3(\Delta^2) = {1 \over 2} [F_u^3(\Delta^2) + F_d^3(\Delta^2)]~.
\label{3ff}
\eq
In the figure, the calculated integral is compared with 
experimental data
of $|F_{ch}^3(\Delta^2)|$ \cite{ffexp}.
It is found that the present approach reproduces well
the data up to a momentum transfer $-\Delta^2=0.25$ GeV$^2$,
which is enough for the aim of this calculation.
In fact, the region of higher momentum transfer 
is not considered here, being
phenomenologically not relevant for the calculation
of GPDs entering coherent DVCS.
The full curve in the
left panel of Fig. 3
is very similar to the one-body 
contribution to the $^3He$ charge f.f. shown in
\cite{ffth}, as it must be, due to the IA used here.
The agreement with data in the relevant kinematical region,
$-\Delta^2 \leq 0.25$ GeV$^2$,
confirms that the inclusion of two-body currents is not
required in the present calculation. 

As an illustration,
the result of the evaluation of 
$H_u^3(x,\xi,\Delta^2)$
and
$H_d^3(x,\xi,\Delta^2)$
by means of Eq. (\ref{spec})
is shown in Figs. 4 and 5, for 
$\Delta^2 =-0.15$ GeV$^2$
and $\Delta^2 =-0.25$ GeV$^2$, respectively,
as a function of $x_3$ and $\xi_3=3 \xi$.
The GPDs are shown
for the $\xi_3$ range allowed by Eq.
(\ref{xim}) and in the $x_3 \geq 0$ region.
The $x_3 \leq 0$ one, being symmetric to the latter,
is not interesting and not shown.

The quality and size of the nuclear effects are discussed
in the next section.

\section{Discussion of nuclear effects}

The full result for the GPD $H_q^3$, Eq. (\ref{spec}),
shown in Figs. 4 and 5, will be now
compared with a prescription
based on the assumptions
that nuclear effects are completely neglected
and the global $\Delta^2$ dependence can be
described 
by
the f.f. of $^3He$:
\bq
H_q^{3,(0)}(x,\xi,\Delta^2) 
= 2 H_q^{3,p}(x,\xi,\Delta^2) + H_q^{3,n}(x,\xi,\Delta^2)~,
\label{app0}
\eq
where the quantity
\bq
H_q^{3,N}(x,\xi,\Delta^2)=  
\tilde H_q^N(x,\xi)
F_q^3 (\Delta^2)
\label{barh}
\eq
represents the flavor $q$ effective GPD of the bound nucleon 
$N=n,p$ in $^3He$. Its $x$ and $\xi$ dependences, given by the function
$\tilde H_q^N(x,\xi)$, 
is the same of the GPD of the free nucleon $N$ (represented 
in this calculation by Eq. (\ref{hdd})),
while its $\Delta^2$ dependence is governed by the
contribution of the quark of flavor $q$ to the
$^3He$ f.f., $F_q^3(\Delta^2)$.

The effect of Fermi motion
and binding can be shown through 
the ratio
\be
R_q^{(0)}(x,\xi,\Delta^2) = { H_q^3(x,\xi,\Delta^2) \over H_q^{3,(0)}
(x,\xi,\Delta^2)} 
\label{rnew}
\eq
i.e. the ratio
of the full result, Eq. (\ref{spec}),
to the approximation Eq. (\ref{app0}).
The latter is evaluated by means of the nucleon GPDs used
as input in the calculation, and taking
\be
F_u^3(\Delta^2) = {10 \over 3} F_{ch}^{3}(\Delta^2)~,
\label{fu3}
\ee 
\be
F_d^3(\Delta^2) = -{4 \over 3} F_{ch}^{3}(\Delta^2)~,
\label{fd3}
\ee
where $F^3_{ch}(\Delta^2)$ is the f.f. which is calculated
within the present approach, by means of Eq. (\ref{ffc}).
The coefficients $10/3$ and $-4/3$ are simply chosen
assuming that the contribution of the
valence quarks of a given flavour to the f.f. of $^3He$
is proportional to their charge. 
One should remember that the normalization of the f.f.
has been chosen in Eq. (\ref{3ff}). 

The choice of calculating the ratio Eq. (\ref{rnew})
to show nuclear effects is a very natural one.
As a matter of fact, the forward limit of the ratio Eq. (\ref{rnew})
is the same of the ratio Eq. (\ref{rat}), yielding the
EMC-like ratio for the parton distribution $q$ and,
if $^3He$ were made of free nucleon at rest,
the ratio Eq. (\ref{rnew}) would be one.
This latter fact can be immediately realized by
observing that the prescription Eq. (\ref{app0})
is exactly obtained by
placing $z=1$, i.e. no Fermi motion effects
and no convolution, into Eq. (\ref{spec}). 
In the present situation the ratio Eq. (\ref{rat})
cannot be used to show nuclear effects, since it does not contain
the $\Delta^2$ dependence coming from the nuclear structure, i.e.
the one provided by the non-diagonal spectral function
in the present approach [cf. Eq. (\ref{spec})].

One should note that
the prescription suggested in Ref. \cite{km}
for finite nuclei in the valence quark sector,
basically assuming that
the nucleus is a system of almost free
nucleons with approximately the same momenta, 
has the same $\Delta^2$ dependence of the 
prescription Eq. (\ref{app0}).


Results are presented in Fig. 6 and 7, where
the ratio Eq. (\ref{rnew}) is shown
for $\Delta^2 = -0.15$ GeV$^2$ 
and $\Delta^2 = -0.25$ GeV$^2$, respectively,
as a function of $x_3$,  
for three different values
of $\xi_3$, for the flavours $u$ and $d$.

Some general trends of the results are apparent:

i) nuclear effects, for $x_3 \leq 0.7$, are as large as 15 \% at most.
Larger effects at higher $x_3$ values are actually related
with the vanishing denominator in Eq. (\ref{rnew}).

ii) Fermi motion and binding have their main effect
for $x_3 \leq 0.3$, at variance with what happens
in the forward limit (cf. Fig. 2).

iii) at fixed $\Delta^2$, nuclear effects increase with
increasing $\xi$, for $x_3 \leq 0.3$.

iv) at fixed $\xi$, nuclear effects increase with increasing
$\Delta^2$, for $x_3 \leq 0.3$.

v) nuclear effects for the $d$ flavour are larger than
for the $u$ flavour.

The behaviour described above can be explained as follows.
As already said in section 2, in IA and in the forward limit, at $x_3=0$
one basically recovers the spectral function normalization
and no nuclear effects, so that the ratio Eq. (\ref{rat}) 
slightly differs from one.
This is not true of course in the present case, due to nuclear
effects hidden not only in the $x'$ dependence, but also
in the $\xi'$ one, according to its definition, Eq. (\ref{xi1}).
Moreover, even if $x_3=\xi_3=0$, in the present situation
the ratio Eq. (\ref{rnew}) does not give the spectral function 
normalization as in the forward case, 
because of the $\Delta^2$ dependence.
One source of such dependence is that,
in the approximation
Eq. (\ref{app0}), it is assumed, through 
Eqs. (\ref{fu3}) and (\ref{fd3}) 
that the quarks $u$ and $d$,
belonging to the protons or to the neutron in
$^3He$, contribute to the charge f.f. in the same way,
being the contribution proportional to their charge only.
Actually, the effect of Fermi motion and binding is stronger
for the quarks belonging to the neutron, having the latter
a larger average momentum with respect to the proton \cite{over,cio}. 
This can be seen
noticing that the pointlike f.f.,  Eq. (\ref{ffp}),
for the proton, shows a stronger $\Delta^2$-dependence, with respect to the
neutron one, the difference being 17 \% (23 \%) at $\Delta^2 = - 0.15$
GeV$^2$ ($\Delta^2 = - 0.25$
GeV$^2$). The prescription given by
Eqs. (\ref{fu3}) and (\ref{fd3}) could be correct only
if the pointlike f.f. had a similar $\Delta^2$ dependence. 
Besides, nuclear effects studied by means
of the ratio Eq. (\ref{rnew}) at fixed $x$ and $\xi$
depend on $\Delta^2$, showing clearly 
that such a dependence cannot be factorized,
i.e. the nuclear GPD cannot be written as the product
of a $\Delta^2$ dependent and a $\Delta^2$ independent term,
confirming what has been found for the deuteron case
in Ref. \cite{cano2}.
One should notice that,
if factorization were valid, Figs. 6 and 7 would be equal.
This fact clearly indicates that a model based on the 
assumption of factorization, such as the one of Ref.
\cite{km}, is not motivated and cannot be used to parametrize
nuclear GPDs for estimates of DVCS cross sections and asymmetries
for light nuclei.

The fact that nuclear effects are larger for the $d$
distribution is also easily explained in terms of
the different contribution of the spectral functions
for the protons and the neutron,
the latter being more important for the GPDs of the $d$ rather than for the
ones of the
$u$ flavour. This has been already shown in the forward
case in Fig. 2, and shortly commented in the previous section.
When the $\Delta^2$-dependence is studied, the effect found
in the forward case
increases, being impossible, as explained above, to really
obtain the $u$ and $d$ contributions to the $^3He$ f.f.
by means of a simple charge rescaling, as it is done in Eqs.
(\ref{fu3}) and (\ref{fd3}).
In particular, the approximation Eq. (\ref{fd3})
for the $d$ flavour is worse than the approximation Eq. (\ref{fu3})
for the $u$ flavour, the disagreement increasing with increasing
$\Delta^2$. One finds that $F_u^3(\Delta^2)$ given by 
Eq. (\ref{fu3}) differs from the one calculated 
through Eq. (\ref{ffc})
by  4 \% (6 \%) at $\Delta^2 = -0.15$ GeV$^2$
($-0.25$ GeV$^2$), while for the $d$ flavour the disagreement 
reaches 9 \% (13 \%) for the same value of the momentum transfer.
This flavour dependence is 
to be expected for any target with isospin different from zero.

As already said, the calculation of DVCS observables 
is beyond the scope of the present paper and will 
be presented elsewhere.
Anyway, a first rough estimate of nuclear effects on 
DVCS observables
can be already sketched.
In fact it is known that 
the point $x=\xi$ gives the bulk of the contribution to 
hard exclusive processes,
since at leading order in QCD the amplitude for DVCS and 
for meson electroproduction just involve GPDs at this point,
which is therefore also the easiest region to access experimentally.
In order to figure out how Fermi motion and binding affect
the ``slice'' $H_q^3(\xi,\xi,\Delta^2)$,
the ratio
(\ref{rnew}) 
for
$x=\xi$, is shown in Fig. 8, as a function
of $\xi_3=x_3$, for the flavor $u$.
It can be seen that even in this crucial region
nuclear effects are systematically underestimated
by the approximation Eq. (\ref{app0}), up to a maximum of 8\%
for the flavor $u$. 

The issue of applying the obtained GPDs to calculate DVCS off $^3He$, to 
estimate cross-sections and to establish
the feasibility of experiments, as it has been done
already for the deuteron target in Ref. \cite{cano2},
is in progress
and will be presented elsewhere.
In particular, the contributions of the incoherent
break-up channel, and of the shadowing effects,
beyond IA, at $\xi \simeq x_{Bj}/2 \leq 0.05$, have to be
evaluated. Besides, the study of polarized GPDs
will be very interesting, due to the peculiar
spin structure of $^3He$ and its implications
for the study of the angular momentum of the free neutron.

\section{Conclusions}

In this paper,
a realistic microscopic calculation
of the unpolarized quark GPD $H_q^3(x,\xi,\Delta^2)$ for $^3He$
is presented.
In an Impulse Approximation framework, a convolution
formula has been obtained where Fermi motion 
and binding effects are properly taken into account 
through an off-diagonal spectral function.
The range of validity of the scheme
is $|\Delta^2| \leq 0.25$ GeV$^2$, the most relevant
for the coherent channel of 
hard exclusive processes.
The forward limit and the
$x$-integral of the obtained
$H_q^3(x,\xi,\Delta^2)$ are formally and numerically
in agreement with the theoretical and experimental knowledge
of the $^3He$ nucleus.

Nuclear effects are found to be larger than in the forward case
and to increase with $\Delta^2$ at fixed
$\xi$, and with $\xi$ at fixed $\Delta^2$.
In particular the latter $\Delta^2$ dependence
does not simply factorize,
in agreement with previous findings for the
deuteron target and at variance
with prescriptions proposed for finite nuclei. 
Moreover, nuclear effects are found to be
larger for the $d$ flavour than for the $u$ one,
being the target non isoscalar.
In particular, it has been shown that
nuclear effects are as large as 8 \%,
even at the crucial
point $x=\xi$.

The shown results represent a prerequisite for
any calculation of DVCS observables.
The evaluation of these quantities, to
establish the feasibility of experiments, 
is in progress and will be presented in a later paper,
together with the contributions of the incoherent,
break-up channel, and of shadowing effects,
beyond IA, at small values of $\xi \simeq x_{Bj}/2$.
The present analysis is also a first step 
towards a complete study of the GPDs for $^3He$,
which could provide, in the polarized case,
with an important tool to unveil details of the
angular momentum structure
of the free neutron.

\acknowledgments

I would like to thank C. Ciofi
degli Atti, L.P. Kaptari, E. Pace
and G. Salm{\`e} for many discussions
about the nuclear three body systems, during our collaboration
in the past years. 
I am also grateful to V. Vento for constant encouragement.

This work is supported in part by MIUR through the funds COFIN03.

\newpage
\appendixonfalse
\section*{Figure Captions}

\vspace{1em}\noindent
{\bf Fig. 1}:
The handbag contribution to the coherent DVCS process
off $^3He$, in the present approach.

\vspace{1em}\noindent
{\bf Fig. 2}:
The ratio Eq. (\ref{rat1})
as a function of $x_3$, for the flavour $u$ (left panel)
and $d$ (right panel).

\vspace{1em}\noindent
{\bf Fig. 3}:
The charge f.f. of $^3He$,
calculated by means of Eq (\ref{3ff})
(full line), is compared with a parametrization
of data \cite{ffexp} (dashed line), in the region of low-$\Delta^2$,
relevant to the present study.
 
\vspace{1em}\noindent
{\bf Fig. 4}:
In the left panel, for the $\xi_3$ values
which are allowed 
at $\Delta^2 = -0.15$ GeV$^2$ according to Eq. (\ref{xim}),
$H_u^3(x_3,\xi_3,\Delta^2)$,
evaluated using Eq. (\ref{main}),
is shown for $0.05 \leq x_3 \leq 0.8$.
The symmetric part at $ x_3 \leq 0$ is not presented.
In the right panel, the same is shown, for the flavour $d$.

\vspace{1em}\noindent
{\bf Fig. 5}:
The same as in Fig. 4,
at $\Delta^2 = -0.25$ GeV$^2$.

\vspace{1em}\noindent
{\bf Fig. 6}:
In the left panel,
the ratio Eq. (\ref{rnew}) is shown, for the $u$ flavour and
$\Delta^2 = -0.15$ GeV$^2$, 
as a function of $x_3$.
The full line has been calculated for $\xi_3=0$,
the dashed line for $\xi_3=0.1$ and the long-dashed
one for $\xi_3=0.2$. 
The symmetric part at $ x_3 \leq 0$ is not presented.
In the right panel, the same is shown, for the flavour $d$

\vspace{1em}\noindent
{\bf Fig. 7}:
The same as in Fig. 6, at
$\Delta^2 = -0.25$ GeV$^2$.



\vspace{1em}\noindent
{\bf Fig. 8}:
The ratio Eq. (\ref{rnew}) 
for the $u$ flavour, for $x_3=\xi_3$,
as a function of $\xi_3$,
at $\Delta^2 = -0.15$ GeV$^2$ (full line),
and at $\Delta^2 = -0.25$ GeV$^2$ (dashed line).


\newpage

\begin{figure}[h]
\vspace{6.6cm}
\includegraphics{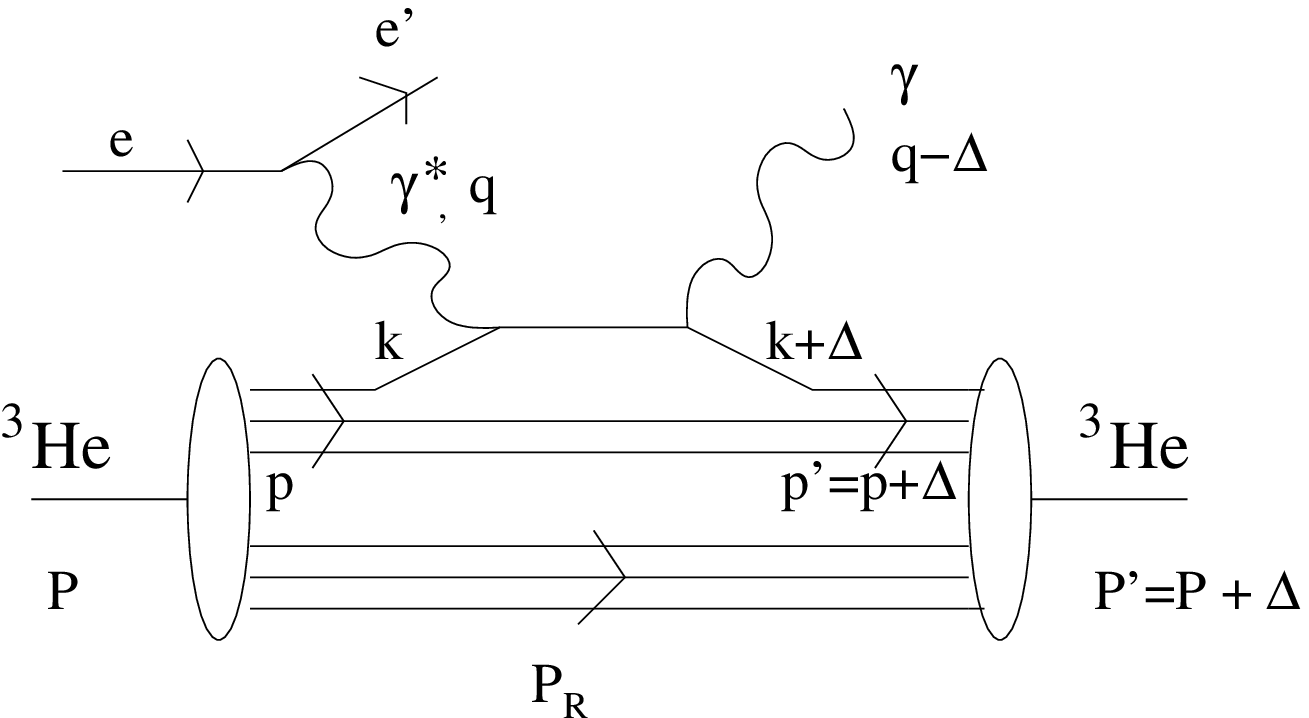}
\caption{}
\end{figure}

\vskip 2cm

\begin{figure}[h]
\vspace{8.6cm}
\includegraphics{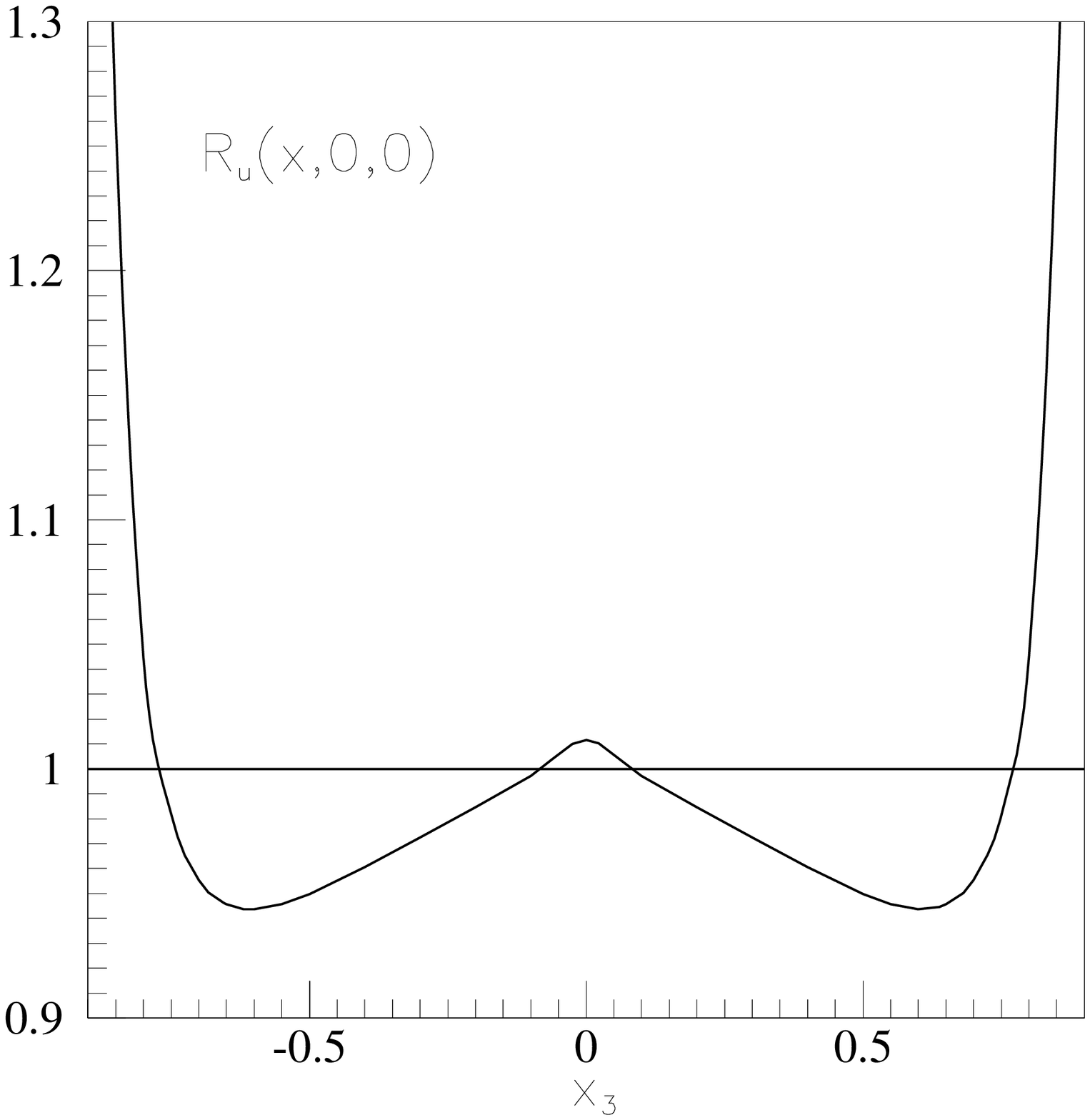}
\includegraphics{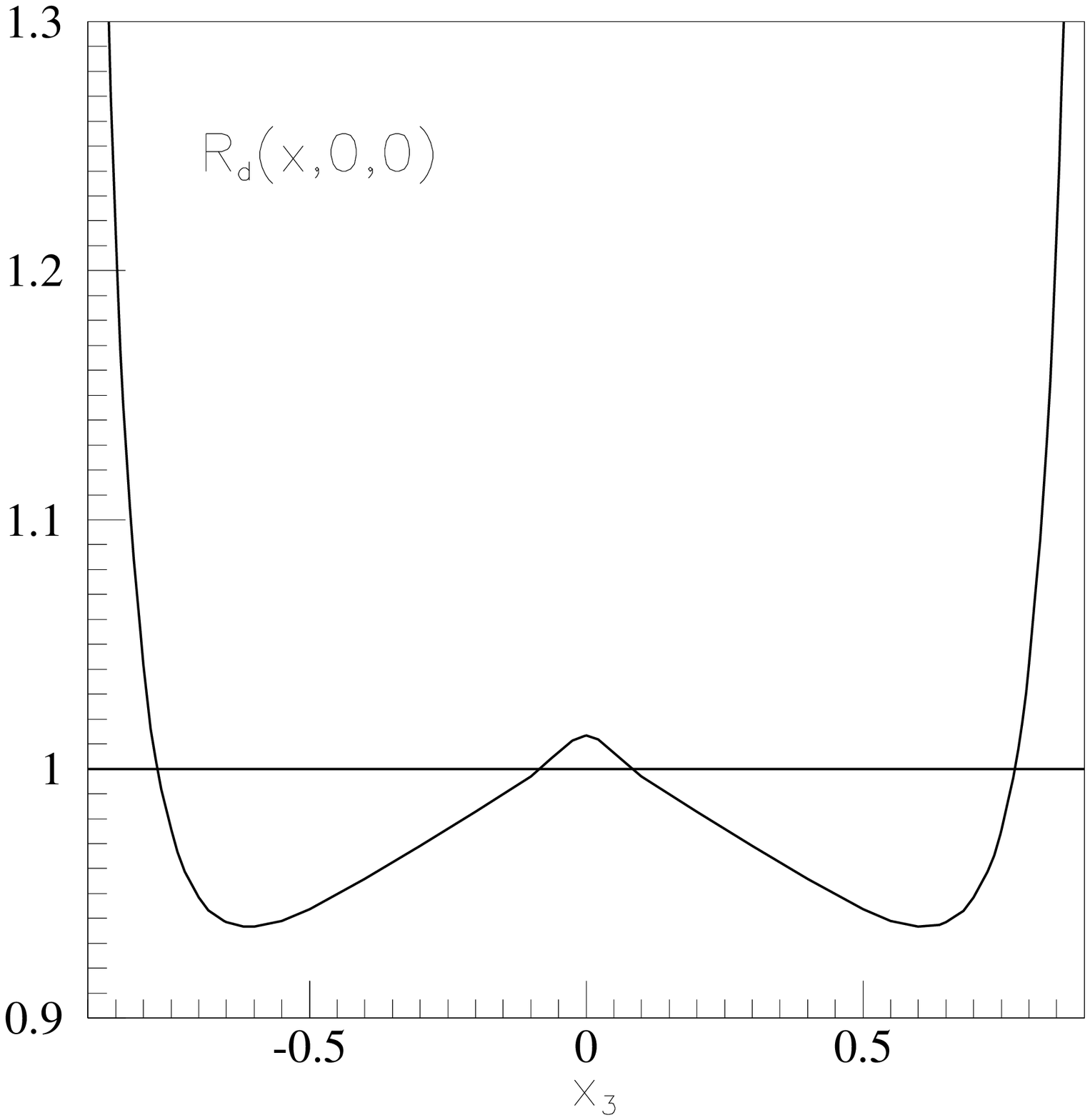}
\caption{}
\end{figure}

\newpage
$ $
\begin{figure}[h]
\vspace{13.0cm}
\includegraphics{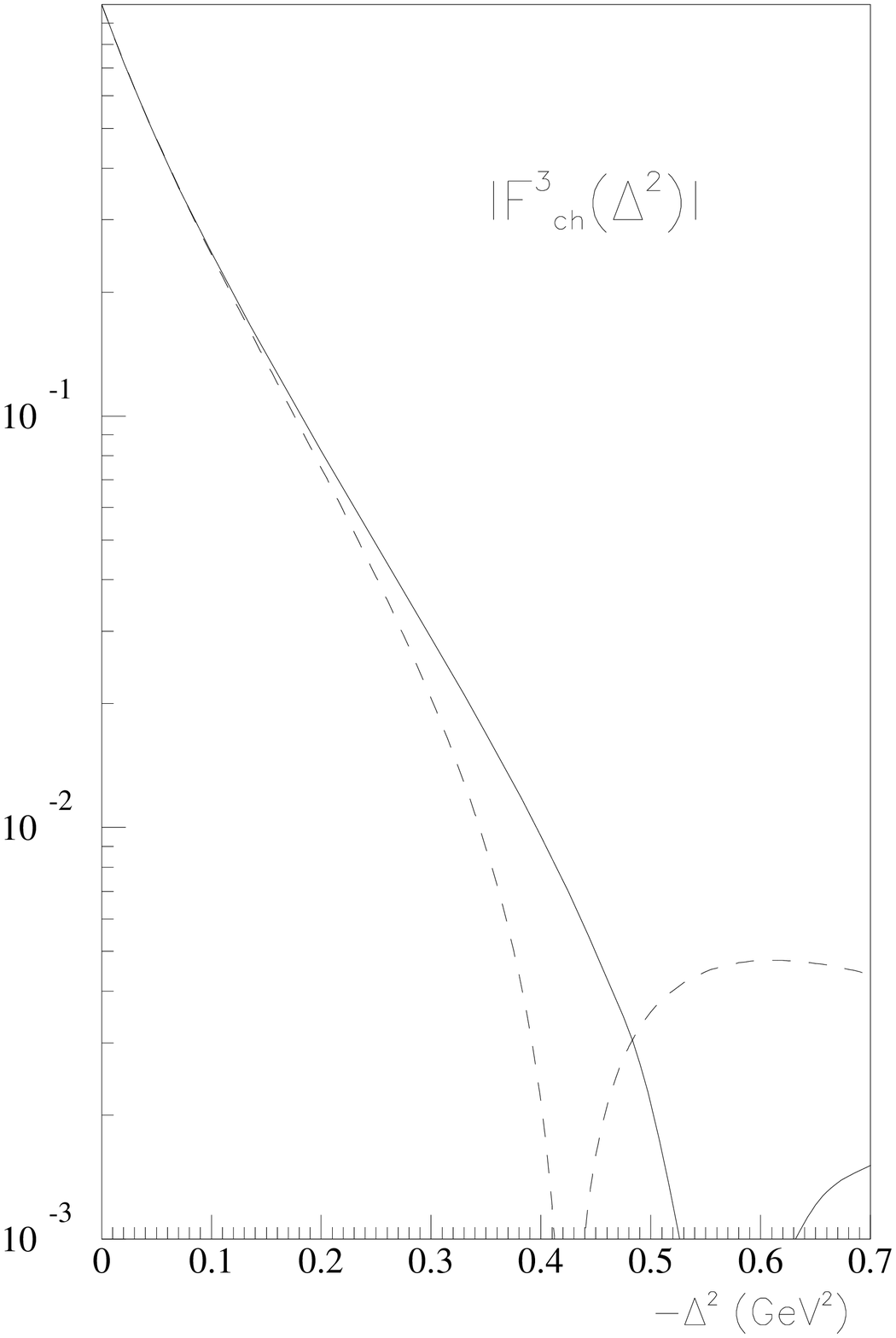}
\caption{}
\end{figure}

\newpage
$ $
\begin{figure}
\vspace{10.0cm}
\includegraphics{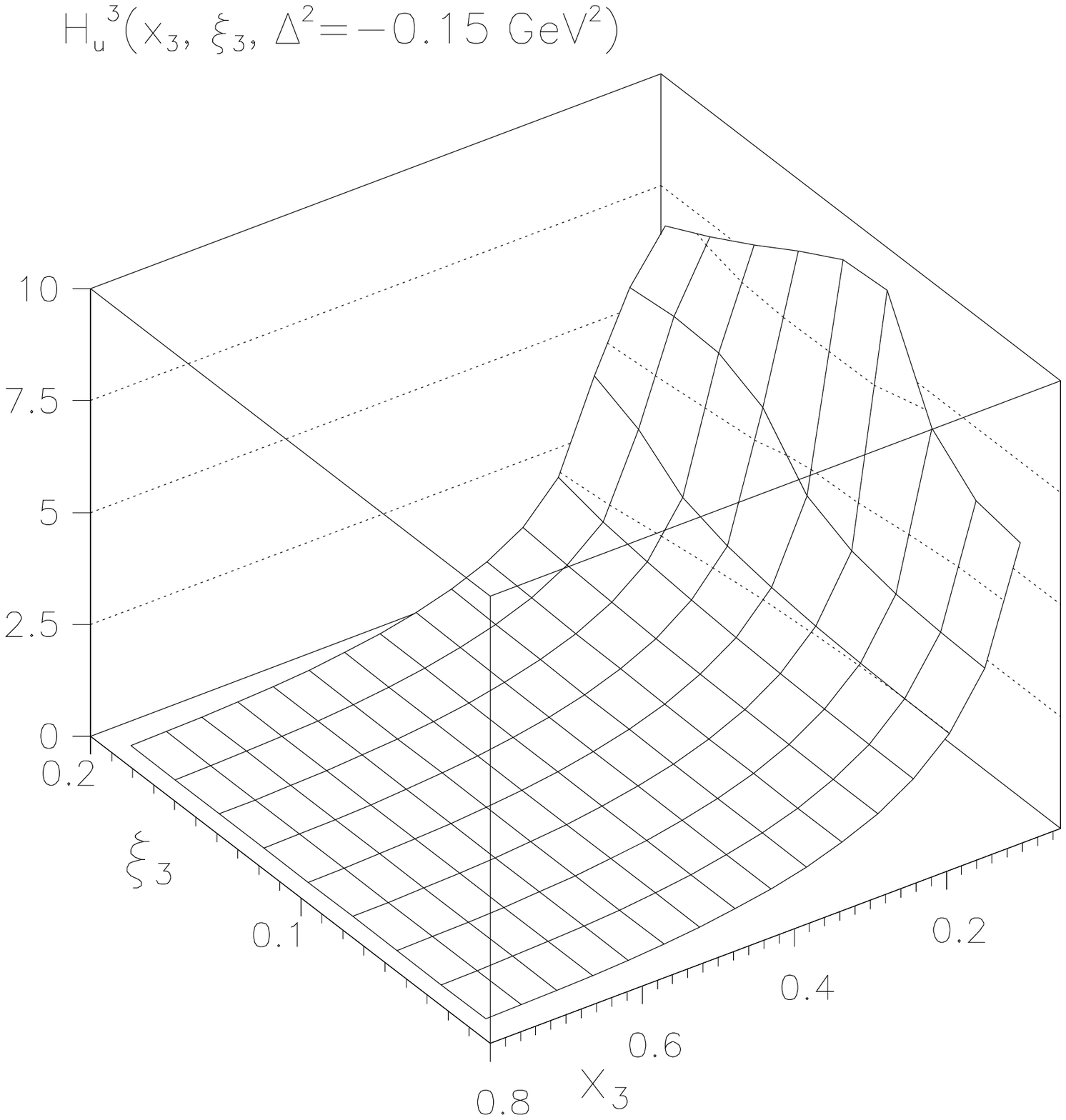}
\includegraphics{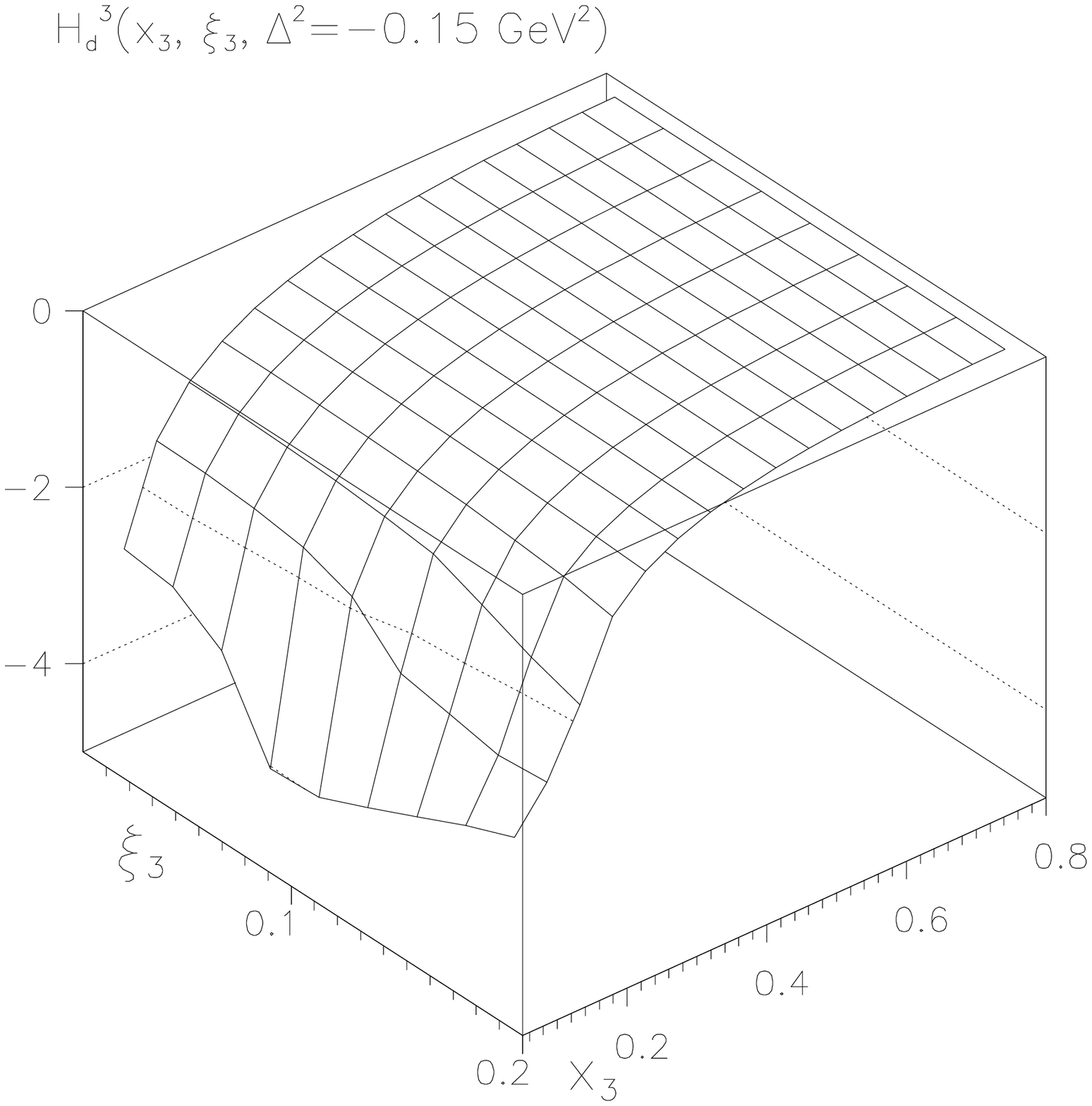}
\caption{}
\vspace{11.cm}
\includegraphics{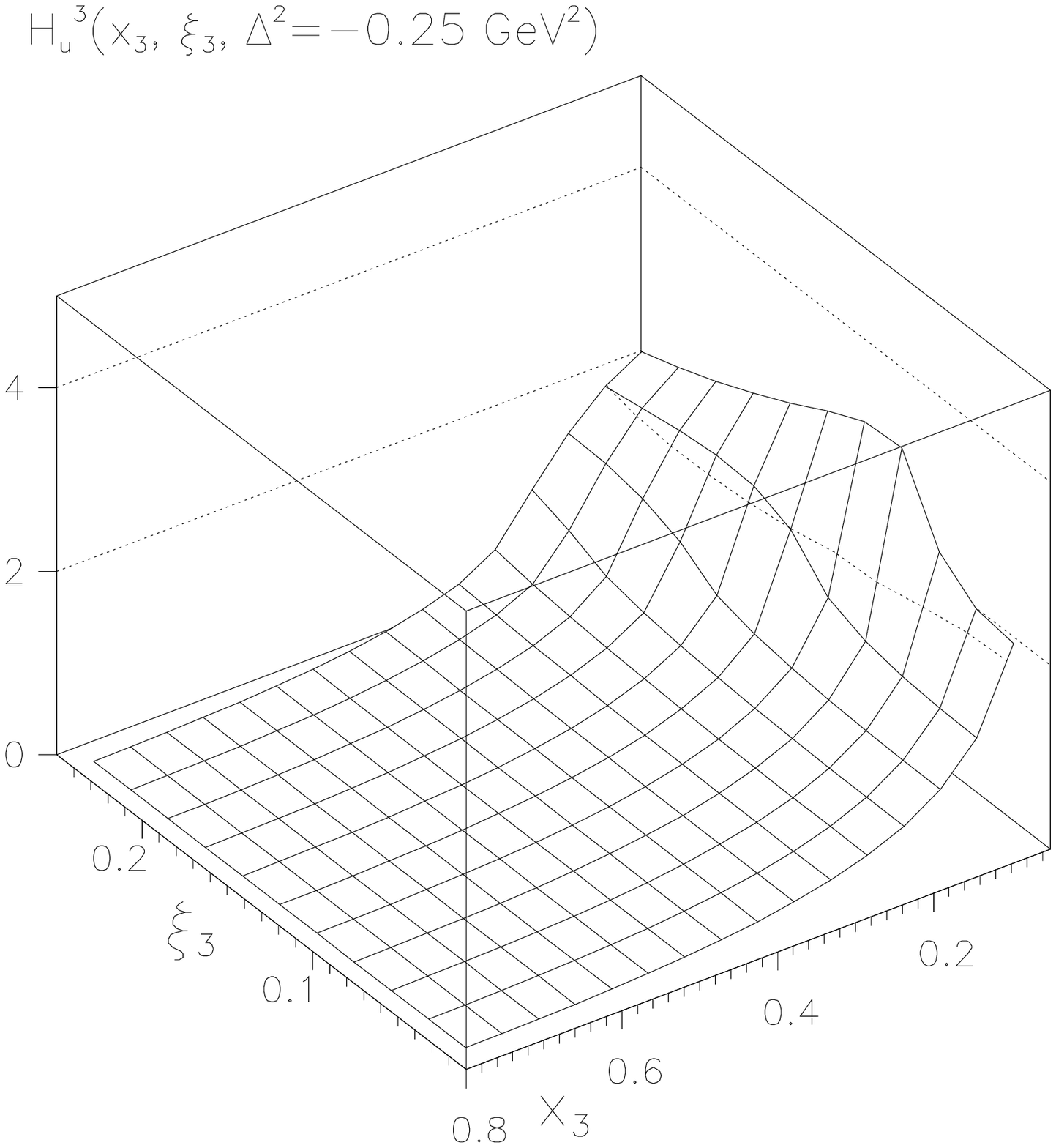}
\includegraphics{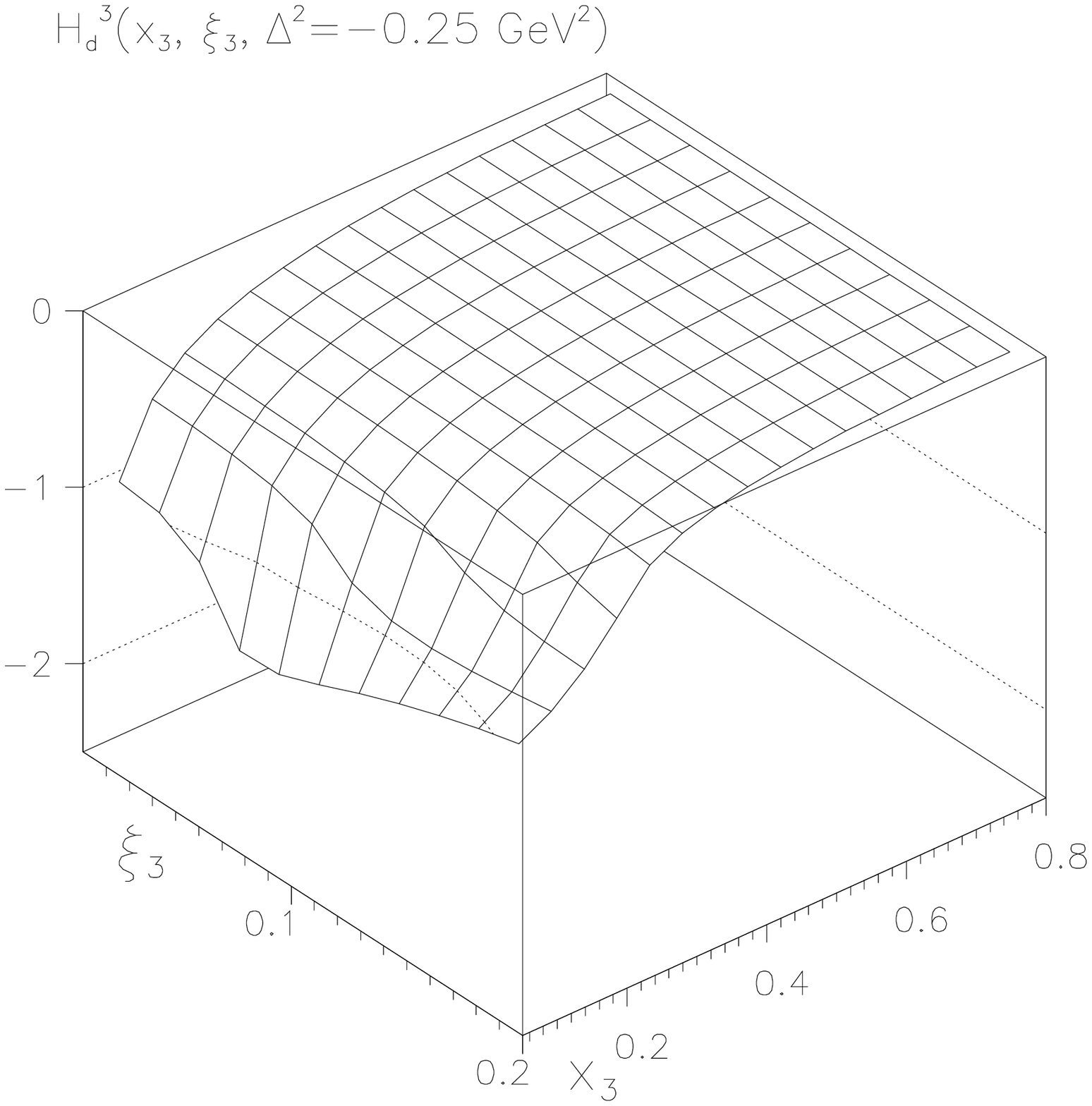}
\caption{}
\end{figure}

\newpage
$ $
\begin{figure}[h]
\vspace{9.2cm}
\includegraphics{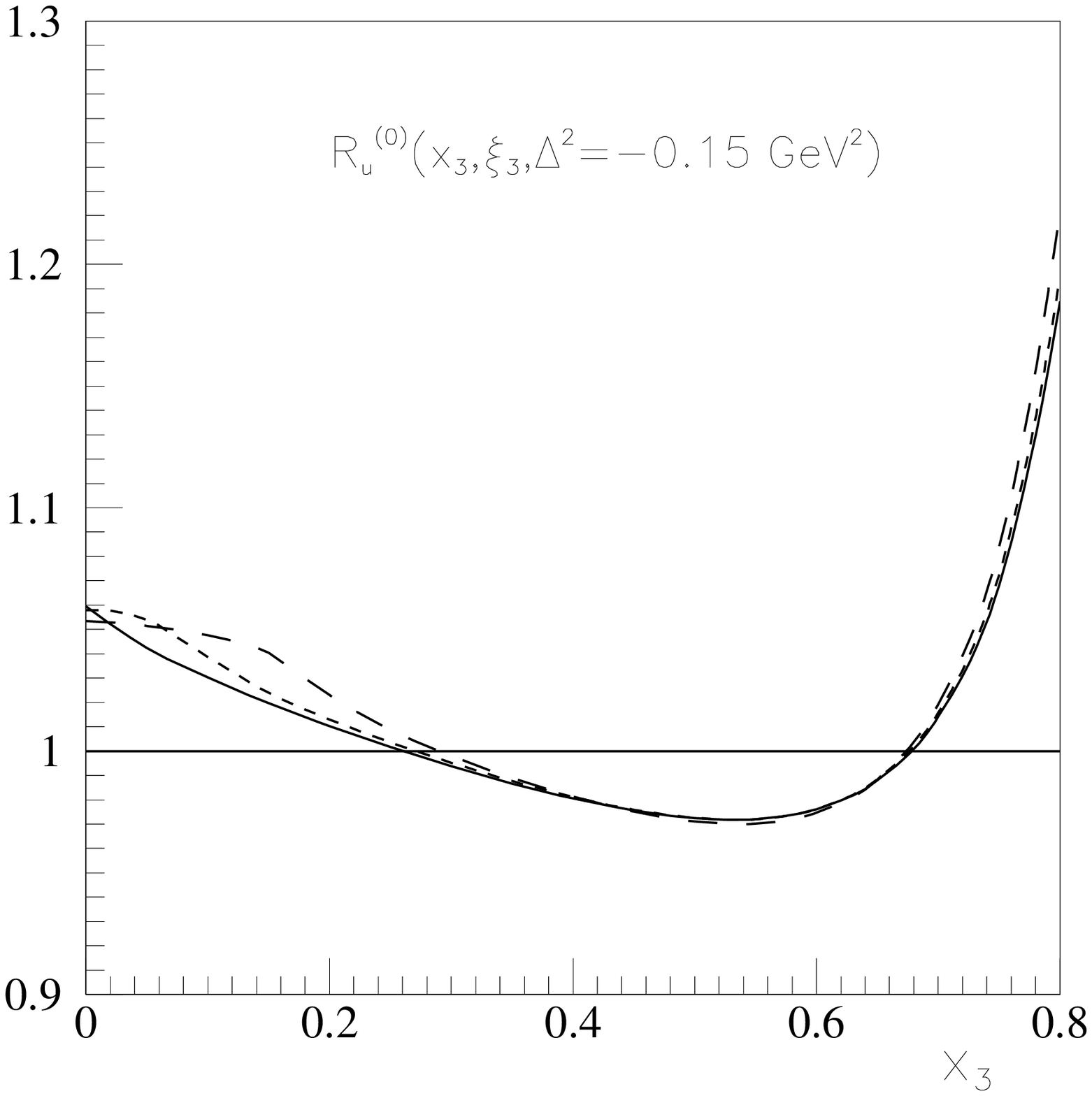}
\includegraphics{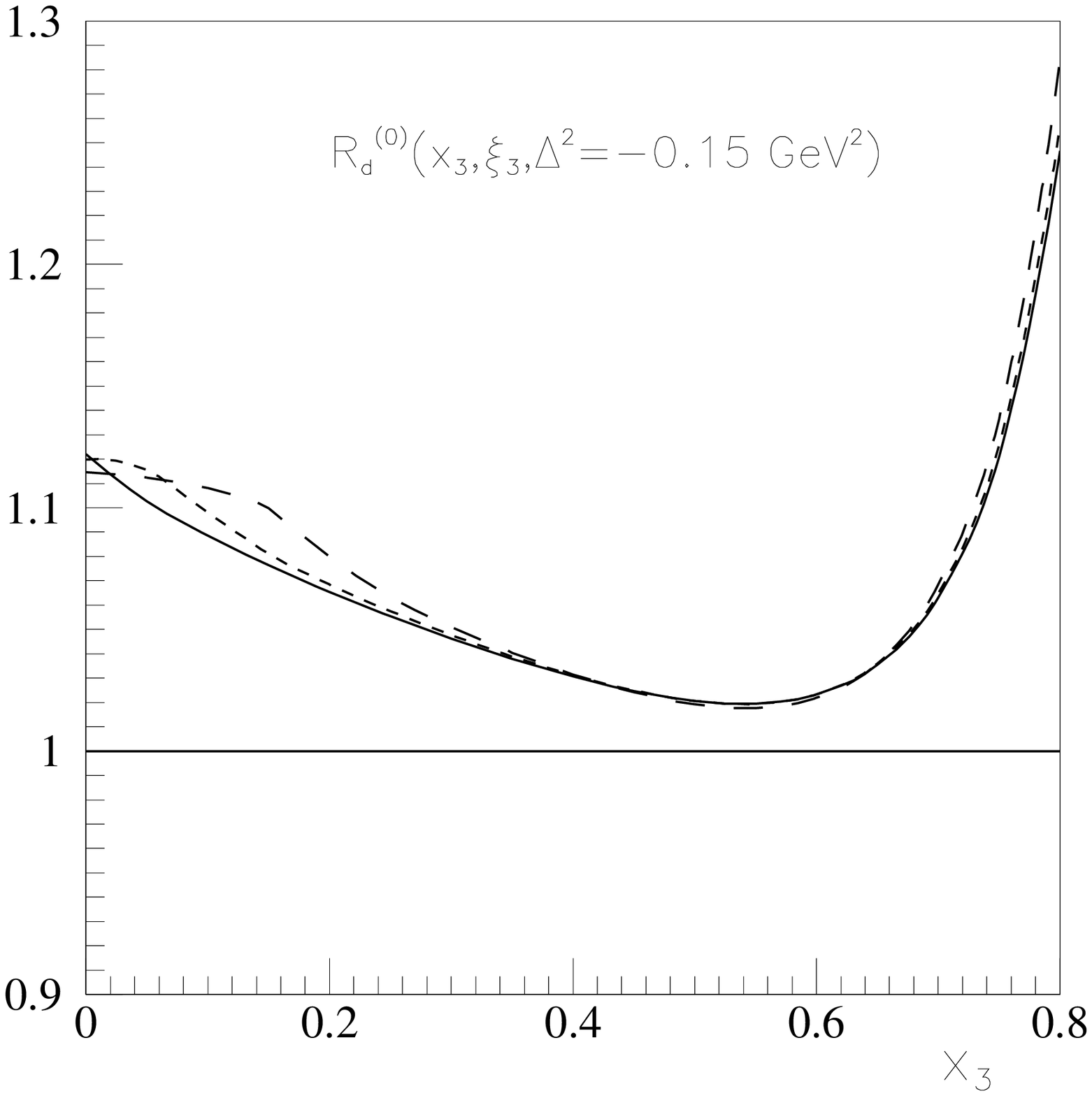}
\caption{}
%
\vskip 1cm
%
\vspace{10.2cm}
\includegraphics{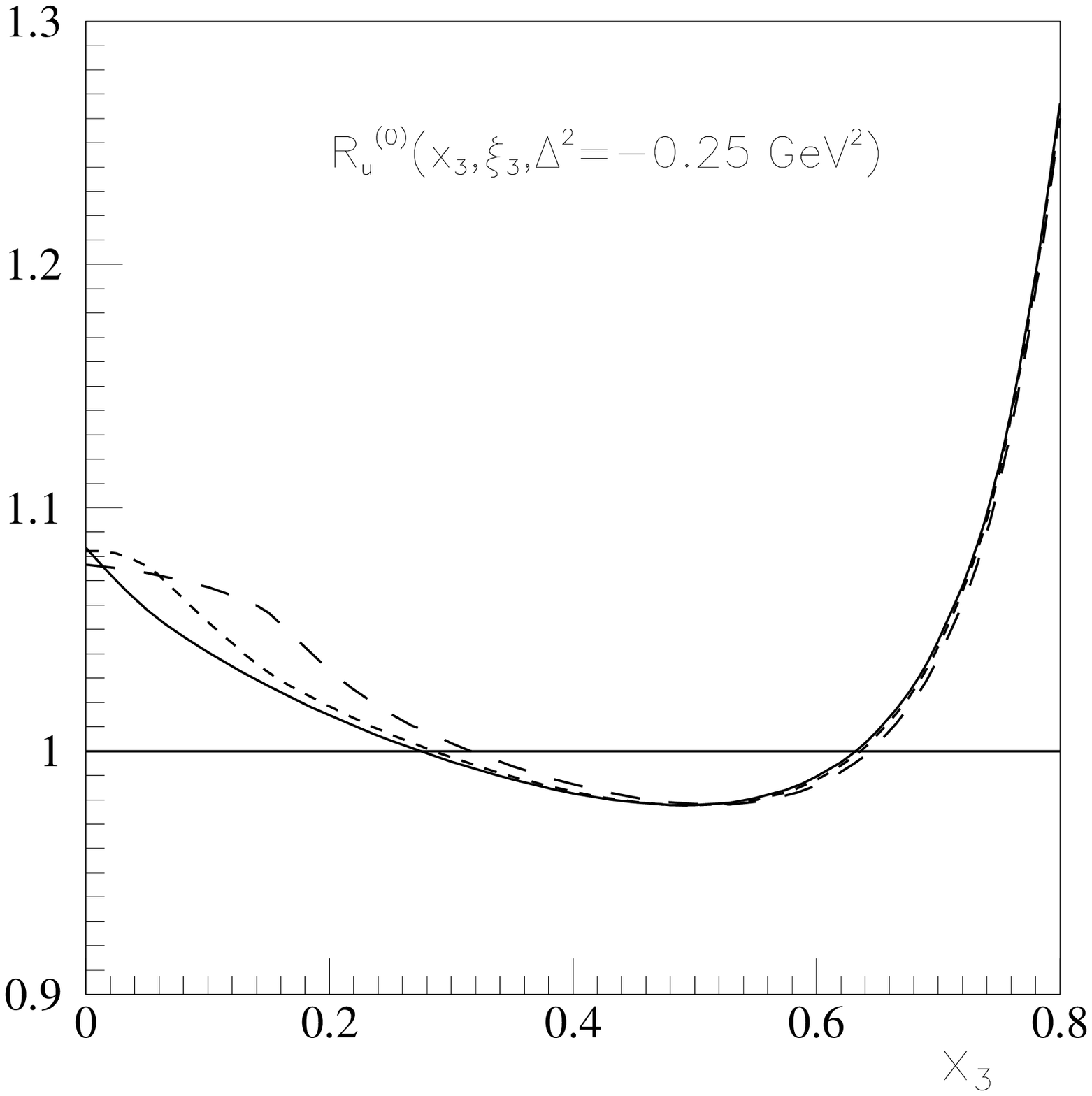}
\includegraphics{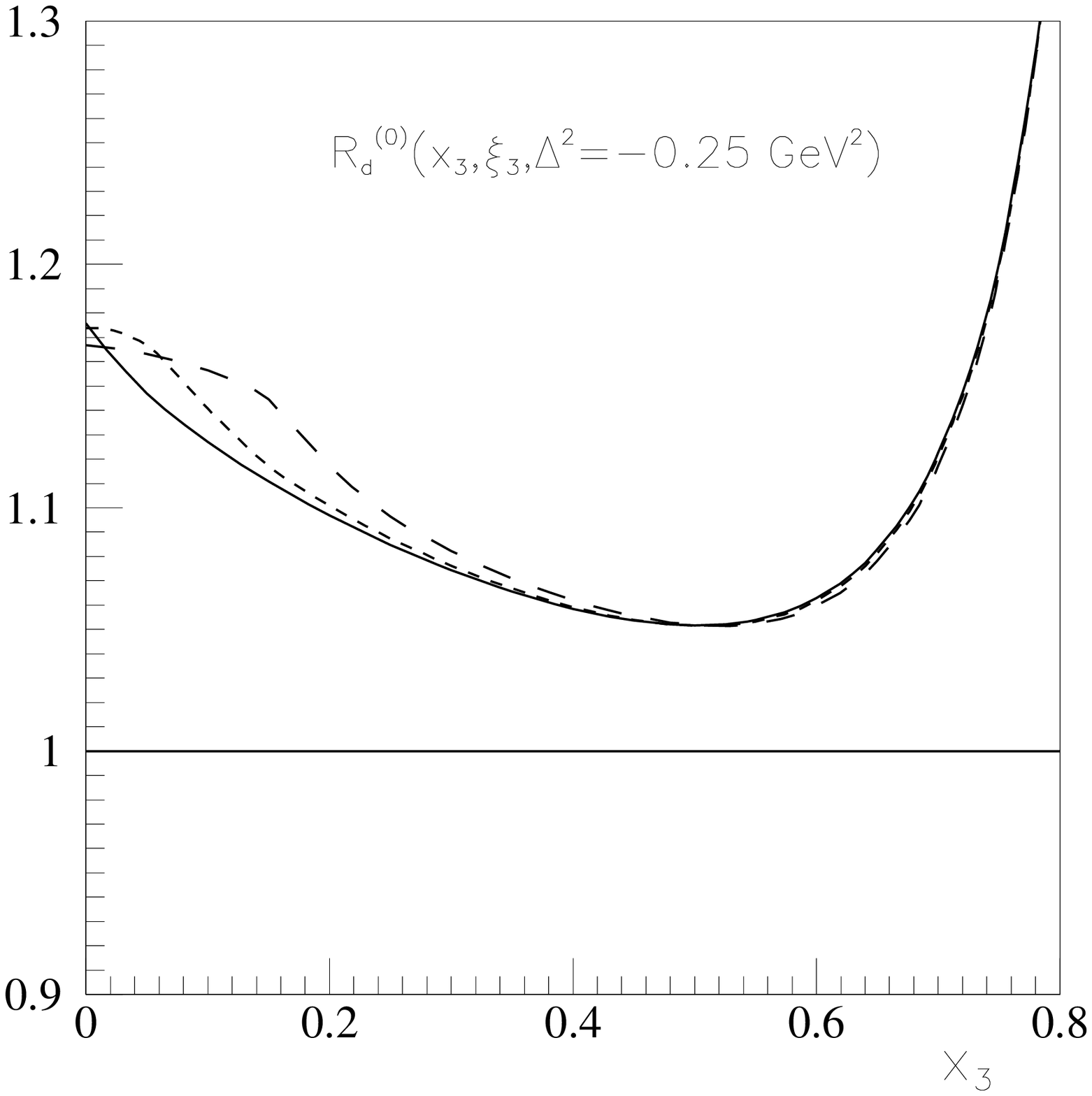}
\caption{}
\end{figure}

\newpage
$ $
\begin{figure}[h]
\vspace{9.2cm}
\includegraphics{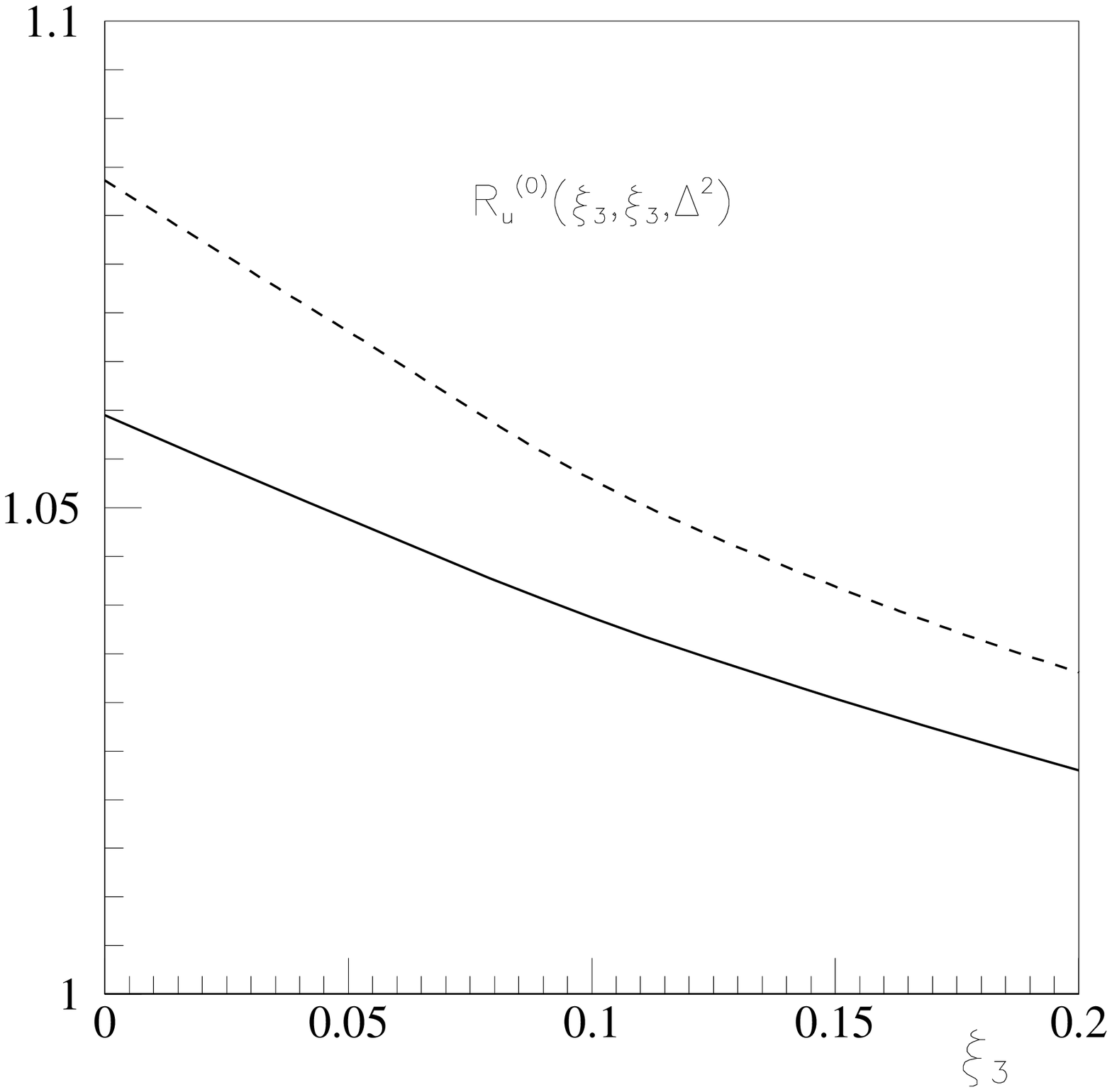}
\caption{}
\end{figure}
%
%


\end{document}